\documentclass[useAMS,usenatbib]{mn2e}

\usepackage{graphicx}
\usepackage{psfig}

\newcommand\aj{AJ}
\newcommand\araa{ARA$\&$A}
\newcommand\apj{ApJ}
\newcommand\apjl{ApJ}
\newcommand\apjs{ApJS}
\newcommand\aap{A$\&$A}
\newcommand\mnras{MNRAS}
\newcommand\prd{Phys.~Rev.~D}
\newcommand\pasp{PASP}
\newcommand\nat{Nature}
\newcommand\fcp{Fund.~Cosmic~Phys.}
\newcommand\physrep{Phys.~Rep.}

\title[Dark matter halo shapes]{Modelling the shapes of the largest gravitationally bound objects}
\author[Graziano Rossi] {Graziano Rossi$^1$\thanks{Email:
    graziano@kias.re.kr}, Ravi K. Sheth$^2$ \& Giuseppe Tormen$^3$\\
\\
$^{1}$ Korea Institute for Advanced Study, Hoegiro 87, Dongdaemun-Gu, Seoul $130-722$, Korea \\
$^{2}$ Center for Particle Cosmology, University of Pennsylvania, 209 South $33^{rd}$ Street, Philadelphia, PA 19104, USA\\
$^{3}$ Dipartimento di Astronomia, Universit\`{a} degli Studi di Padova, Vicolo dell'Osservatorio, 2 I-35122 Padova, Italy}  
\date{Accepted ?? Received ?? ; in original form ??}
\date{\today}
\pagerange{\pageref{firstpage}--\pageref{lastpage}}
\pubyear{2010}

\begin{document}
\maketitle
\label{firstpage}



\def\gnl{g_{\rm NL}}
\def\fnl{f_{\rm NL}}
\def\be{\begin{equation}}
\def\ee{\end{equation}}
\def\bea{\begin{eqnarray}}
\def\eea{\end{eqnarray}}
\def\bn{\begin{enumerate}}
\def\en{\end{enumerate}}
\def\nn{\nonumber}
\def\pa{{\partial}}
\def\l{\left}
\def\r{\right}



\begin{abstract}
We combine the physics of the ellipsoidal collapse model with the 
excursion set theory to study the shapes of dark matter halos.  
In particular, we develop an analytic approximation to the 
nonlinear evolution that is more accurate than the Zeldovich 
approximation;
we introduce a planar representation of halo axis ratios, which 
allows a concise and intuitive description of the dynamics of 
collapsing regions and allows one to relate the final shape of 
a halo to its initial shape; 
we provide simple physical explanations for some empirical fitting 
formulae obtained from numerical studies.
Comparison with simulations is challenging, as there is no 
agreement about how to define a non-spherical gravitationally 
bound object. 
Nevertheless, we find that our model matches the conditional 
minor-to-intermediate axis ratio distribution
rather well, although it 
disagrees with the numerical results in reproducing the 
minor-to-major axis ratio distribution.  In particular, the 
mass dependence of the minor-to-major axis distribution appears to 
be the opposite to what is found in many previous numerical studies, 
where low-mass halos are preferentially more spherical than high-mass 
halos. In our model, the high-mass halos are predicted to be more 
spherical, consistent with results based on a more recent and elaborate 
halo finding algorithm, and with observations of the mass dependence of 
the shapes of early-type galaxies.  We suggest that 
some of the disagreement with some previous numerical studies may 
be alleviated if we consider only isolated halos. 
\end{abstract}






\section{Introduction} \label{intro}


A snapshot of a high-resolution dark matter simulation,  
at relatively low redshift, reveals that halos are neither
spherically symmetric nor smooth.  This is because, without even 
considering all the small scale complications arising from baryonic 
physics (i.e. pressure effects, merging, cooling, heating), 
the statistics of a Gaussian random field implies that spherically 
symmetric initial configurations should be a set of measure zero 
(Doroshkevich 1970). 
While it has long been recognized that spherical collapse is too 
idealized to be realistic, and that the true gravitational process 
must, at the very least, be ellipsoidal (Icke 1973; 
White \& Silk 1979; Barrow \& Silk 1981; Kuhlman et al. 1996), 
most analytic models of structure formation make the simplifying 
assumption that gravitationally bound objects are spherical, and 
formed from a spherical collapse 
(i.e. Gunn \& Gott 1972; Lahav et al. 1991; Lacey \& Cole 1993; 
Lokas \& Hoffman 2001). 

It has also become common practice to identify halos in simulations 
using a spherical overdensity algorithm, which finds the mass around 
isolated peaks in the density field such that the mean interior density 
is $\Delta$ times the background density, where the value of
$\Delta$ is motivated by the spherical top-hat model (Lacey \& Cole 1994). 
This is despite the fact that dark matter halos that form in simulations 
are rather elongated, and, in general, strongly triaxial:  close to prolate 
(minor and intermediate axes are comparable in size to each other, 
and much smaller than the biggest axis) in the central parts, and
rounder in the outskirts (Barnes \& Efstathiou 1987; Frenk et
al. 1988; Dubinski \& Carlberg 1991; Katz 1991; Warren et al. 1992;
Jing et al. 1995; Thomas et al. 1998; Jing \& Suto 2002; 
Allgood et al. 2006; Bett et al. 2007; Diemand, Kuhlen \& Madau 2007;
Hayashi et al. 2007; Kuhlen et al. 2007; Mu{\~n}oz-Cuartas et
al. 2010; Wang et al. 2010).
This shape variation leads to a significant variance in local 
properties, compared to the spherically averaged value at a given
radius, that are now becoming of interest.  


Studying and quantifying the degree of halo triaxiality is 
of broader importance.
In fact, in the current paradigm of hierarchical clustering, 
dark matter halos are the hosts within which gas cools and collapses 
to form galaxies (White \& Rees 1978; White \& Frenk 1991), thus 
making them the building blocks of the large scale structure (LSS) 
of the Universe (Cooray \& Sheth 2002).  
Hence, understanding the assembly histories, kinematics, clustering,
and fundamental structural properties of halos -- such as their 
intrinsic shapes -- is the first necessary step in understanding 
the properties of galaxies (Mo, Mao \& White 1998; Dutton et al. 2007; 
Diemand et al. 2008). In turn, the formation of dark matter halos 
affects the properties of the galaxies hosted by the halos; therefore,
inspection of the galaxy distribution in redshift surveys such as the
SDSS (York et al. 2000) allows one to relate the properties of
galaxies to those of their host halos.  
Moreover, the characteristic density of a halo appears to track the 
mean density of the Universe at the time of its formation 
(e.g. Zhao et al. 2009), leading to a quasi-universal profile 
(Navarro, Frenk \& White 1996; Kormendy \& Freeman 2004), although self-similarity is 
not preserved: different halos cannot be rescaled to look alike 
(Navarro et al. 2004, 2010; Merritt et al. 2006; Gao et al. 2008; 
Reed et al. 2010).
 

Triaxiality also has a number of observationally relevant implications.
For example, modeling of dark matter halos beyond the spherical 
approximation is crucial in understanding the nonlinear clustering of
halos and dark matter (Sheth, Mo \& Tormen 2001), the formation and 
evolution of galaxies (Cole \& Lacey 1996), and their relation to the 
cosmic web (Shen et al. 2006; Hahn et al. 2007; 
Lee, Hahn \& Porciani 2009a,b; Forero-Romero et al. 2009; 
Pogosyan et al. 2009; Sousbie et al. 2009; Park, Kim \& Park 2010).  
In particular, the higher order statistics of the nonlinear density 
field is sensitive to halo triaxiality (Smith \& Watts 2005; 
Smith, Watts \& Sheth 2006).  

Triaxiality represents a useful framework for the non-spherical 
modelling of the intracluster gas, which recent observations suggest 
will be key in deriving more accurate temperature profiles of X-ray 
clusters, and in general for cosmological parameter determinations 
via the Sunyaev-Zeldovich effect (Lee et al. 2005).  For example, 
Kawahara (2010) has derived the axis ratio distribution of X-ray 
clusters in the XMM-Newton catalog of Snowden et al. (2008) and 
confirmed that the typical X-ray halo is well approximated by a 
triaxial ellipsoid.  
And recently, Morandi, Pedersen \& Limousin (2010) have presented the 
first determination of the intrinsic triaxial shapes and three-dimensional 
physical parameters of both dark matter and the intra-cluster medium 
for the galaxy cluster Abell 1689.  

Triaxiality is also useful in predicting -- and hence can be constrained 
by -- a variety of gravitational lensing observations, including weak 
and strong lens statistics (Bartelmann 1995; 
Van Waerbeke et al. 2000; Schulz et al. 2005; Brada{\v c} et al. 2006; 
Carbone et al. 2006; Bernstein 2007; Riquelme \& Spergel 2007; 
Broadhurst et al. 2008; Limousin et al. 2008; 
Schneider \& Er 2008; Mandelbaum et al. 2008, 2009; 
Zitrin et al. 2009; Bernstein \& Nakajima 2009; J{\"o}nsson et al. 2010), 
and gravitational flexion (e.g., Hawken \& Bridle 2009).  
By measuring the shapes of dark matter halos, galaxy-galaxy lensing 
can provide constraints on galaxy formation models and the nature of 
dark matter (Hoekstra et al. 2004; Mandelbaum et al. 2006; 
Parker et al. 2007). 

Surveys like ESA's Euclid mission (http://sci.esa.int/euclid) will in fact provide 
accurate data for shape estimates through ``cosmic shear'', a direct 
measure of the metric fluctuations in the Universe 
(Hoekstra \& Jain 2008; Bernstein 2010; Rhodes et al. 2010), 
which in turn constrain dark energy properties (Albrecht et al. 2006). 
A primary source of noise in such measurements is due to the difficulty 
in distinguishing between intrinsic galaxy shapes and shape distortion 
due to lensing (Bartelmann \& Schneider 2001; Refregier 2003; 
Hoekstra et al. 2005; Mandelbaum et al 2006; Bridle et al. 2009).  
Hence accurate modelling of the correlated shapes and orientations 
dark matter halos can be extremely useful.
The higher order statistics of the nonlinear density field in such 
surveys is also sensitive to halo triaxiality (Smith et al. 2006).  

On galaxy mass scales, an understanding of halo triaxiality provides
useful input to studies of galactic disks in triaxial halos.  
E.g., Jeon, Kim \& Ann (2009) considered the fundamental dynamics 
between the disk and the axisymmetric or triaxial halo, and 
Valluri et al. (2010) analyzed the orbital structure of dark 
matter particles in $N$-body simulations in an effort to understand 
what is the physical mechanism driving shape changes caused by 
growing central masses (also see Debattista et al. 2008).  
This shape change reconciles the strongly prolate-triaxial shapes found 
in collisionless $N$-body simulations with observations, which  generally
find much rounder halos (see for example Banerjee \& Jog 2008).  
Finally, resolving the fine grained structure of galaxy mass halos enables 
one to make more realistic predictions for direct and indirect dark matter 
detection experiments (e.g. Giocoli, Pieri \& Tormen 2008).


The shapes of dark matter halos can be quantified using high-resolution 
$N$-body simulations of hierarchical gravitational clustering. This is 
currently the best way to address many of the tasks above.  
Simulations are becoming increasingly accurate in mass and spatial 
resolution:  e.g., 
the Millennium Run (Springel et al. 2005), 
the Horizon Run (Kim et al. 2009), 
the Millennium II (Boylan-Kolchin et al. 2009) 
and the Bolshoi Simulation (Klypin et al. 2010).

In these numerical studies, 
many different aspects have been considered over a wide
range of physical scales and cosmic histories.
Common findings suggest that the total mass is the key in 
determining the final shapes of halos, although other environmental
parameters may play a role in the process. In general, halos do 
exhibit a rich variety of shapes with a preference for prolateness 
over oblateness.
More massive halos tend to be less spherical and more prolate, 
and are preferentially aligned with primordial filaments, 
while less massive halos are in general rounder 
(e.g., Jing \& Suto 2002; Allgood et al. 2006; Araya-Melo et al. 2009).  
On the other hand, perturbation theory suggests that the most massive 
objects should be spherical (Bernardeau 1994; Pogosyan et al. 1998), and Lemson (1995) 
showed that the spherical model does provide a good description 
of the evolution of the spherically averaged profile.  
More recently, Dalal et al. (2008) reported that massive halos are 
indeed very well described by the spherical collapse model.  This was 
recently confirmed by Park et al. (2010), who provided a clear 
explanation for the discrepancy with previous measurements.  

The shapes of subhalos are similar to those of host halos, but subhalos 
tend to be a bit rounder, especially the ones near the host 
halo center. Tidal interactions make individual subhalos rounder over time
and they tend to align their major axis towards the center of the host
halo. Formation of halos is also affected by the large-scale environment,
which may have an impact on their shapes, and those shapes can be modified 
by galaxy formation as well.  Subhalos are not the subject of our study.  
 

There are many more numerical studies of halo shapes than analytic 
models.  This is because the formation, evolution and virialization 
of dark matter halos is complex; no rigorous analytic techniques are 
available for use in both the linear and the nonlinear regimes. In 
addition, choosing the appropriate definition of halo shapes is subtle.  
For example, Eisenstein \& Loeb (1995) describe an analysis of halo 
shapes which uses the ellipsoidal collapse model of Bond \& Myers (1996).  
However, as we discuss below, their definition of a halo differs from 
the more commonly accepted definition.  Moreover, they present results 
for collapsed objects that had the same initial density.  Since the 
time it takes for to collapse is a complicated function of density 
{\em and} shape (Bond \& Myers 1996), this means that they compare 
objects of one shape at one time with those of another shape at a 
different time.  This is rarely measured in simulations:  the shape 
distribution of most interest is, of course, that for a fixed time 
(e.g., halos at $z=0$).  


The main purpose of the present work is to provide a simple model for 
the distribution of halo shapes at any given time as seen in the 
simulations, starting from first principles.  Following Rossi (2008), 
our analytic prescription has two independent parts: 
the first is a scheme for how an initially spherical patch evolves and
virializes; the second is the correct assignment of initial shapes to 
halos of different masses.  

In this respect, our model is similar in philosophy to that of 
Lee, Jing \& Suto (2005).  However, there are important differences.  
(1) They assume the Zeldovich approximation remains valid even during 
the nonlinear regime, where it is known to fail; 
(2) They assume a spherical collapse threshold for the formation 
of halos, or an empirical recipe based on Lee \& Shandarin (1998).
In contrast, because we are modelling triaxial objects, we 
self-consistently use ellipsoidal, rather than spherical collapse 
dynamics to generate our predictions.  

To describe the evolution of non-spherical structures we adopt the 
ellipsoidal collapse model of Bond \& Myers (1996), which was used 
by Sheth, Mo \& Tormen (2001) to estimate of how the abundance of 
dark matter halos depends on halo mass. 
In this model, dark matter halos are identified with ellipsoids which 
have collapsed completely along all three axes (we show below that, 
in effect, the Eisenstein \& Loeb 1995 definition corresponds to 
collapse along just two axes).  In this framework, the time required 
to collapse depends on the overdensity $\delta_{\rm i}$ and size
$R_{\rm i}$ 
of the initial patch, and on the surrounding shear field, parametrized 
by its ellipticity $e$ and prolateness $p$.  Requiring the collapse to 
happen at a given time makes $\delta_{\rm i}$ a function of $e$ and $p$ 
(Sheth et al. 2001):  $\delta_{\rm ec}(e,p)$.  The combination of 
$\delta_{\rm i}=\delta_{\rm ec}$, $e$ and $p$ determines the axis ratios of 
the object at all times, and, in particular, at the final time.  
Thus ($e,p$) establishes the time of collapse (it was this fact which 
was exploited by Sheth et al. 2001), as well as the axis ratios at 
collapse (a fact we exploit here).

The second part is the correct assignment of ($e,p$) values to halos 
of different masses.  We do this following Sheth \& Tormen (2002) 
(also see Chiueh \& Lee 2001; Sandvik et al. 2007).  
In essence, the correct ($e,p$) distribution is specified by the
statistics of Gaussian random fields.  In a Gaussian random field,
the distribution of ($e,p$) values depends on the size and overdensity 
of the patch: $g(e, p|\delta, R)$.  
Massive halos form from larger patches in the initial conditions 
than do less massive halos, so we expect the distribution of initial 
($e,p$) values, and hence the distribution of final axis ratios, to 
also depend on halo mass. Thus, although the generic evolution is 
initially towards an oblate, pancake-like structure, followed by a 
shift towards a more prolate shape, as the other axes also begin to 
shrink, quantifying this evolution, and merging it with the correct 
(mass dependent) initial distribution of shapes is the main focus of 
this work.


The outline of this paper is as follows. 
In Section \ref{nl_dynamics} we present the ellipsoidal model for the
gravitational collapse of an initially spherical patch; we discuss a 
reasonably accurate analytic approximation to the evolution, more 
rigorous than the Zeldovich approximation;  
we introduce the axis ratio plane, which provides a concise description 
of the dynamics of collapsing regions and allows one to relate the 
final shape of a halo with its original, pre-collapsed, shape.
In Section \ref{initial_conditions} 
we present
the full model for halo shapes,
we explain how the initial 
conditions are obtained via the excursion set algorithm, and we expand 
on the prolateness distribution -- crucial in understanding our main results.
In Section \ref{simulation_comparison} 
the model is contrasted with high resolution $N$-body simulations. 
A simple explanation of an empirical relation found by Jing \& Suto (2002) is
given, and a number of caveats in the comparison model/simulations are 
highlighted.  
Finally, Section \ref{discussion} discusses in detail limitations in 
the modelling and difficulties arising from numerical studies, and 
suggests future improvements. 


In our calculations we assume a spatially 
flat cosmological model with 
$(\Omega_{\rm M}, \Omega_{\rm \Lambda}, h) = (0.3, 0.7, 0.7)$, where
$\Omega_{\rm M}$ and $\Omega_{\rm \Lambda}$ are the present day densities 
of matter and cosmological constant scaled to the critical density. 
We write the Hubble constant as $H_0 = 100h$ km s$^{-1}$ Mpc$^{-1}$.
Regarding the overall notation, we always use the subscript $i$ to denote
initial quantities, and the subscript $f$ for final quantities. 
The expansion factor of the Universe, with mean density $\bar{\rho}$,
is represented by the small case letter $a$ -- not to be confused
with the capital letter $A$, which denotes the axis ratios of an 
ellipsoidal patch.



\section{Nonlinear dynamics} \label{nl_dynamics}

The nonlinear gravitational evolution of a medium is complex, even 
without complications arising from gas dynamics. 
It involves smooth accretion, tidal interactions with the environment 
as well as violent episodes of collisions with other halos, merging 
and fragmentation. 
In what follows, we first briefly summarize the key aspects of the
ellipsoidal collapse. We then discuss an analytic approximation for 
the evolution, and introduce a planar representation 
of halo axis ratios; this allows for a concise description
of the dynamics of collapsing regions, and provides a mapping between 
the initial and final shape of a halo.


\subsection{A model for the gravitational collapse} \label{ecollapse_model}


The simplest description of the gravitational evolution and
virialization of a cosmic structure is the spherical collapse model 
(Gunn \& Gott 1972; Gott 1975; Gunn 1977; Fillmore \& Goldreich 1984;
Bertschinger 1985; Mo \& White 1996). 
In this framework, an isolated overdensity in an otherwise unperturbed 
universe first expands with the Hubble flow, then turns around and 
collapses. However, the initial shear field, rather than the density, 
has been shown to play a crucial role in the formation of nonlinear 
structures (Zeldovich 1970; Hoffman 1986, 1988; Peebles 1990; 
Dubinski 1992; van de Weygaert \& Babul 1994; Audit \& Alimi 1996; 
Audit, Teyssier \& Alimi 1997). 
Therefore, refinements to the spherical approximation which include 
local shear effects can be obtained by introducing an ellipsoidal 
collapse scheme
(Bond \& Myers 1996; Eisensein \& Loeb 1995; Monaco 1995, 1997, 1998; 
Lee \& Shandarin 1998; Chiueh \& Lee 2001; Sheth, Mo \& Tormen 2001; 
Sheth \& Tormen 2002; Ohta et al. 2004; Shen et al. 2006; 
Sandvik et al. 2007; Desjacques 2008; Desjacques \& Smith 2008; Rossi 2008). 

In this study we adopt the homogeneous ellipsoidal model in the form of
Bond \& Myers (1996), although the ellipsoidal collapse has a long 
history (Lin, Mestel \& Shu 1965; Icke 1973; White \& Silk 1979; 
Barrow \& Silk 1981).
To lowest order, their algorithm reduces to Zeldovich's (1970) 
approximation, and so linear theory is reproduced.  
In this framework, an initially spherical patch of initial size
$R_{\rm i}$ 
with overdensity $\delta_{\rm i}$ is distorted by the shear field into a 
collapsing homogeneous ellipsoid.  
The exterior tidal force arising from the matter outside of the
ellipsoid is completely determined by the volume-averaged strain of
the ellipsoid.  The details of the substructure in the interior are 
ignored, and the strongly nonlinear internal dynamics of the collapsing 
patch are assumed to be largely decoupled from the weakly nonlinear 
dynamics describing the motion of the patch itself; in this respect, 
the homogeneous ellipsoid picture may be thought of as a tensor virial 
theorem approach to the average interior dynamics. 
In fact, in the nonlinear regime the one-to-one correspondence between 
the external tidal field and the local strain tensor is no longer true 
on small scales where the rms density fluctuation $\sigma \gg 1$; 
therefore, one would not expect the simple ellipsoidal model to apply 
in this regime.

There is another sense in which this approach is only a simple 
approximation.  
It assumes that the inertia tensor of the final bound object is 
perfectly correlated with the local tidal tensor in the initial 
Lagrangian space.
Measurements in simulations show that the two tensors are not 
perfectly correlated (e.g. Lee \& Pen 2000).  On the other hand, 
the correlation is stronger than naive tidal torque theories predict 
(e.g., Porciani, Dekel \& Hoffman 2002), so the assumption of perfect 
correlation is a useful idealization.   

It is usual to characterize the initial shear field by the ellipticity 
$e$ and prolateness $p$ associated with the potential rather than with 
the density field (i.e. Bardeen et al. 1986).  
This is because the components of the $3\times 3$ strain tensor 
are the second derivatives of the potential.  The eigenvalues of the 
initial strain tensor are related to the initial density contrast and 
to the shear ellipticity and prolateness by:
\begin{equation}
\lambda_1(t_{\rm i}) = {\delta(t_{\rm i}) \over 3} (1 + 3e + p),
\label{lambda1}
\end{equation}
\begin{equation}
\lambda_2(t_{\rm i}) = {\delta(t_{\rm i}) \over 3} (1 -2p),
\label{lambda2}
\end{equation}
\begin{equation}
\lambda_3(t_{\rm i}) = {\delta(t_{\rm i}) \over 3} (1 - 3e + p).
\label{lambda3}
\end{equation}
Note that $\sum_{\rm j} \lambda_{\rm j} = \delta_{\rm i}$.  
If $\delta > 0$, then $e \ge 0$ and $-e \le p \le e$, so 
$\lambda_1 \ge \lambda_2 \ge \lambda_3$.

\begin{figure}
\begin{center}
\includegraphics[angle=0,width=0.48\textwidth]{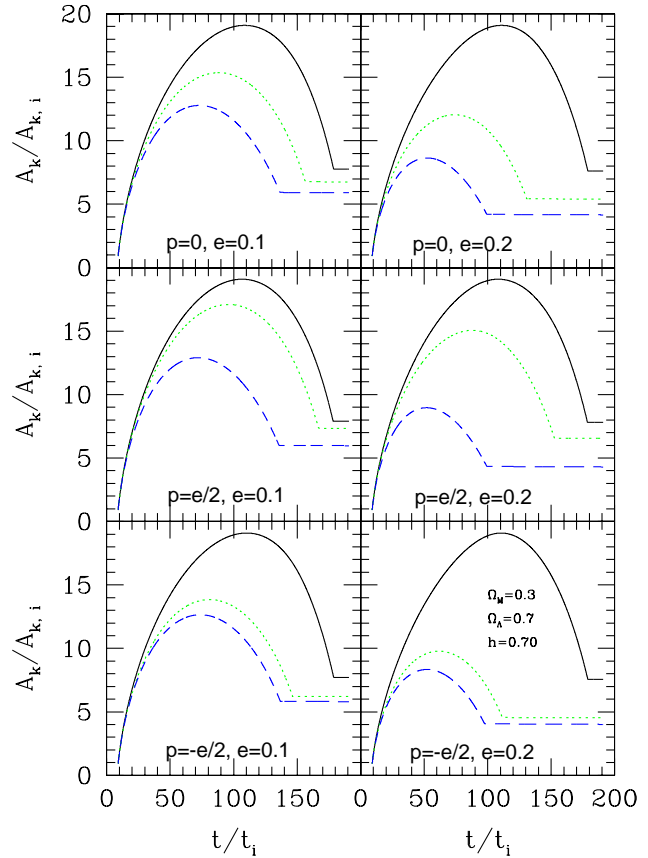}   
\caption{Evolution of axis lengths in our model, in physical units. 
         The times at which different axes freeze out are determined
         by the initial ($e,p,\delta$) values, as specified in the panels.}
\label{ellipsoid_evolution_fig}
\end{center}
\end{figure}

If we denote with $A_{\rm k}$ the scale factors for the three principal axes
of the ellipsoid, then the initial conditions are set by the Zeldovich 
approximation, both for the displacement and the velocity fields:
\begin{equation}
A_{\rm k}(t_{\rm i}) = a(t_{\rm i}) [1-\lambda_{\rm k}(t_{\rm i})],
\label{initial_axes}
\end{equation}
\begin{equation}
{{\rm d}A_{\rm k}(t_{\rm i}) \over {\rm d}t} = H(t_{\rm i})\Big [ A_{\rm
  k}(t_{\rm i}) - a(t_{\rm i}) \lambda_{\rm k}(t_{\rm i}) {{\rm d~ln} D  \over
  {\rm d~ln} a } \Big |_{\rm t \equiv t_{\rm i}} \Big ]. 
\label{initial_velocities}
\end{equation}
Notice that $A_3 \ge A_2 \ge A_1$.  The subsequent evolution is given by 
\begin{equation}
\frac{{\rm d}^2 A_{\rm k}}{{\rm d}t^2} = \Omega_{\rm \Lambda} H_{0}^2
 A_{\rm k} - 4 \pi G
\bar{\rho} A_{\rm k} \Big[\frac{1+\delta}{3}+\frac{b'_{\rm k}
    \delta}{2}+\lambda'_{\rm ext,k} \Big],
\label{ellipsoid_evolution}
\end{equation}
\noindent where $\bar{\rho}$ is the mean density of the Universe,
$\delta$ the relative overdensity, and $b'_{\rm k}=b_{\rm k}-2/3$ and
$\lambda'_{\rm ext,k}$ account for the interior and exterior tidal
forces.  In particular,
\begin{equation}
b_{\rm k}(t)= \Big[ \prod_{\rm m=1}^{\rm 3} A_{\rm m}(t) \Big ]
\int_{0}^{\infty} { {\rm d}\tau \over [A_{\rm k}^2(t)+\tau] \prod_{\rm
    j=1}^3 [A_{\rm j}^2(t)+\tau]^{\rm 1/2}}
\label{bk} 
\end{equation}
and 
\begin{equation}
\lambda'_{\rm ext, k} (t) = \frac{D(t)}{D(t_{\rm i})}
  \Big[\lambda_{\rm k}(t_{\rm i})-\delta(t_{\rm i})/3 \Big],
\label{linear_ext_tide_approx}
\end{equation}
with $D$ the linear theory growing mode.
Other possibilities for $\lambda'_{\rm ext,k}$ include the 
`nonlinear' (Bond \& Myers 1996), or the 
`hybrid' (Angrick \& Bartelmann 2010) approximations.  
However, in our case we are concerned with later times, where a 
different choice for the external shear field does not significantly 
affect our conclusions.  

If $e=p=0$, then all three eigenvalues equal $\delta_{\rm i}/3$, hence all three 
axes have the same length initially.  In this case,  the exterior anisotropic 
tidal force is zero and $b'_{\rm k}=0$, so one gets the usual cycloid
solution for a closed universe.  Hence, all three axes evolve similarly, 
so the object remains spherical, and the time to collapse is determined 
by one number:  $\delta_{\rm i}$.  
But for more general initial values of $\delta(t_{\rm i})$, $e$ and $p$, 
and an initial redshift, equation (\ref{ellipsoid_evolution}) must be solved
numerically for each axis $A_{\rm k}$.  Generically, a triaxial object 
has three critical times, corresponding to the collapse along each of 
the three axes.  In this case, the shortest axis, $A_1$, collapses 
first and $A_3$ collapses last.  

Figure \ref{ellipsoid_evolution_fig} shows how the lengths of the 
three axes evolve in the model, in physical units. In all cases 
we set $\delta(t_{\rm i})= [D(t_{\rm i})/D(t_{\rm 0})] \delta(t_0)$, 
with $\delta(t_0)=1.6753$ being the usual critical value associated 
with spherical collapse in the adopted cosmology at $z=0$. 
Our numerical calculations start at a time $t_{\rm i}$, which 
corresponds to a redshift $z_{\rm i}=39$.  Each panel show results 
for a different pair of ($e,p$).
For a given $e$, $p>0$ implies a pancake-like structure (i.e. one 
short axis and two long), while $p<0$ results in filament-like structures. 
The main point of the figure is to illustrate that, in this model, 
a given pair ($e,p$) determines the axis ratios of the object at all 
times, and, in particular, at the final time.


Halos are identified with objects that have collapsed along all three axes. 
Bond \& Myers (1996) stop collapse along axis $k$ by simply freezing
$A_{\rm k}$ once a critical radius $A_{\rm eq,k} = a\, f_{\rm r}$ is reached
during the infall phase, where $a$ is the expansion factor of the
Universe and typically the radial freeze-out factor $f_{\rm r} = 0.177$
is chosen in order to reproduce the `virial' density contrast of 
$\Delta = a^3/(A_1A_2A_3)=179$ familiar from spherical top-hat 
calculations in an Einstein-de Sitter model; $f_r$ must be computed for 
more complicated cosmologies.

In our study, we slightly modify the stop criterion proposed by 
Bond \& Myers (1996) as follows:
we still freeze the value of axis $i$ once a critical radius is 
reached during the infall phase (the radial freeze-out factor being 
chosen in order to reproduce the correct spherical virial density
contrast in the assumed cosmological model), but we progress the 
evolution in time till we reach the point in which each axis has 
collapsed completely, in order to consider fully relaxed halos.  
Bond \& Myers (1996) found that the time at which the longest axis 
of the ellipsoid freezes out is relatively insensitive to the 
exact value of $f_{\rm r}$, within a given cosmological model.  
In this respect, a more sophisticated collapse criterion such as 
the one recently proposed by Angrik \& Bartelmann (2010), and based 
on the tensor-virial theorem, does not affect our analysis.

However, this stop criterion is rather different from that used by 
Eisenstein \& Loeb (1995), and this difference does matter.  
In their prescription, the collapse is stopped at the time when a 
sphere, whose overdensity was that of the initial ellipsoid, would 
have shrunk to zero radius. This happens to be very close to the time 
when the \textit{intermediate} axis collapses (Shen et al. 2006, and 
see discussion below), so it can be substantially before the time 
that the third axis collapses.  

Before moving on, we note that the critical density required for collapse 
by the present time is well-approximated by
\begin{equation}   
 \label{deltaec_approx}
 \delta_{\rm ec} (e,p) \simeq \frac{\delta_{\rm sc}}{1-\beta \sqrt{5(e^2 \pm p^2)}}
\end{equation}
with $\beta=0.365$.  This will be useful in what follows.  


\begin{figure}
\begin{center}
\includegraphics[angle=0,width=0.50\textwidth]{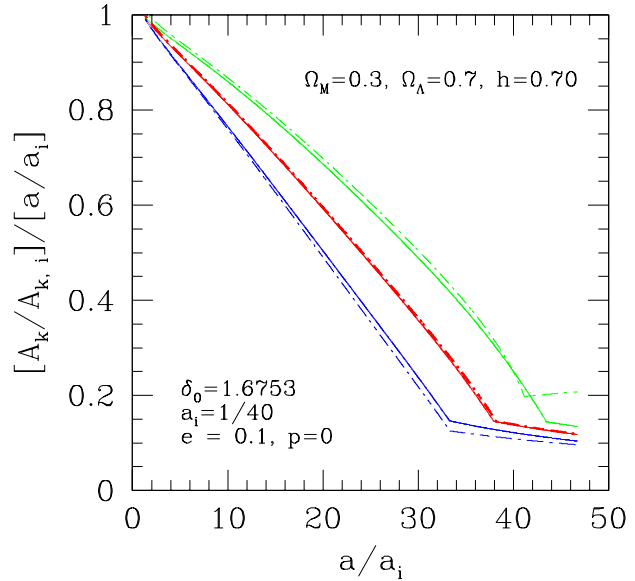}   
\caption{Comoving evolution of axis lengths in our ellipsoidal model. 
   Solid lines are obtained by solving equation (\ref{ellipsoid_evolution}) 
   numerically; dashed-dotted lines show equation~(\ref{WS_approx}). 
 }
  \label{ellipsoid_evolution_approx_fig}
\end{center}
\end{figure}

\begin{figure*}
\begin{center}
\includegraphics[angle=0,width=0.95\textwidth]{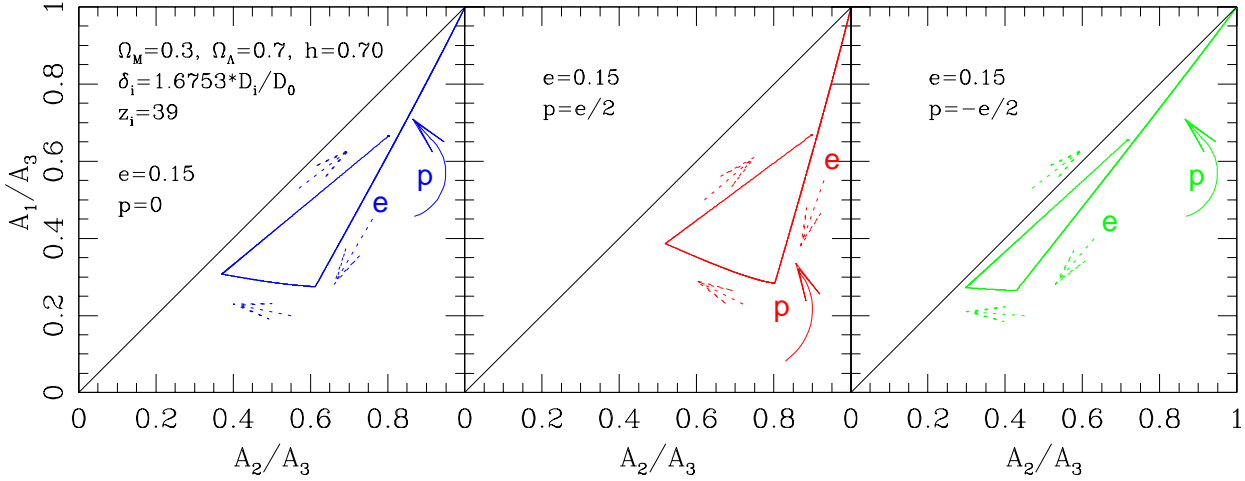}  
\caption{The `axis ratio plane' for dark matter halos.
  Left, central and right panels show $p=0$ and $p=\pm~e/2$,
  respectively, when $e=0.15$. The initial redshift is $z_{\rm i}=39$, for an
  initially spherical patch with overdensity
  $\delta(t_{\rm i})= [D(t_{\rm i})/D(t_{\rm 0})] \delta(t_0)$, 
  where $\delta(t_0)=1.6753$.  See the main text for more details.}
\label{ax_rat_plane}
\end{center}
\end{figure*}


\subsection{An analytic approximation} \label{analytic_proxy}

To characterize the dynamics of a collapsing object, one must 
solve numerically the coupled partial differential equations for the
ellipsoidal collapse, as shown in Figure \ref{ellipsoid_evolution_fig}.
However, there is a useful analytic approximation to the exact solution 
which provides considerable insight.  By extending results in 
White \& Silk (1979), Shen et al. (2006) show that 
\begin{eqnarray}
{A_{\rm k} (t) \over A_{\rm k} (t_{\rm i})}  \,\frac{a(t_{\rm i})}{a(t)}
 &\simeq& 1-\frac{D(t)}{D(t_{\rm i})}
  \lambda_{\rm k}(t_{\rm i})  \nonumber \\ & & - {A_{\rm h}(t_{\rm i}) \over
  A_{\rm k}(t_{\rm i})}\Big[1-\frac{D(t)}{D(t_{\rm i})}{\delta(t_{\rm i})\over 3}
  - \frac{a_{\rm e}(t)}{a(t)}\Big],
\label{WS_approx}
\end{eqnarray}
where
 $A_{\rm h}(t_{\rm i})=3/\sum_{\rm j} A_{\rm  j}^{-1}(t_{\rm i})$ and
 $a_{\rm e}(t)$ is the expansion factor of a universe with initial 
density contrast
 $\delta(t_{\rm i}) \equiv \sum_{\rm j} \lambda_{\rm j}(t_{\rm i})$.  
This approximation to the full ellipsoidal collapse model can be 
thought of as correcting the Zeldovich (1970) approximation by the 
factor by which it is wrong for a sphere (Lam \& Sheth 2008).

To first order in $\delta(t_{\rm i})$, 
\begin{equation}
\frac{A_{1,3}(t)}{A_{1,3}(t_{\rm i})} \frac{a(t_{\rm i})}{a(t)}
  \simeq \frac{a_{\rm e}(t)}{a(t)} - \frac{\delta(t_{\rm i})}{3} (p\pm 3e)
  \Big [1 + {D(t) \over D(t_{\rm i})} - {a_{\rm e} (t) \over a(t)} \Big ]
\label{a13_proxy_2nd}
\end{equation}
\begin{equation}
\frac{A_2(t)}{A_2(t_{\rm i})} \frac{a(t_{\rm i})}{a(t)}   \simeq
\frac{a_{\rm e}(t)}{a(t)} + \frac{\delta(t_{\rm i})}{3} 2p\,\Big [
  1 + {D(t) \over D(t_{\rm i})} - \frac{a_{\rm e} (t)}{a(t)} \Big ].
\label{a2_proxy_2nd}
\end{equation}
Notice that when $p=0$, then 
\begin{equation}
\frac{A_2(t)}{A_2(t_{\rm i})} \simeq \frac{a_{\rm e}(t)}{a(t_{\rm i})};
\label{a2_proxy}
\end{equation}
in this case, the second axis evolves exactly as in the spherical
model, since for the spherical case 
$e \equiv p  = 0$, $\lambda_{\rm
  k}(t_{\rm i}) \equiv \lambda(t_{\rm i}) \equiv \delta(t_{\rm
  i})/3$ and so $A_{\rm k}(t) \equiv A(t) \rightarrow A(t_i)
[a_e(t)/a(t_i)]$. 
In general, the second axis evolves very similarly to that for a 
spherical model with the same initial overdensity $\delta(t_{\rm i})$.  
In this respect, a spherical collapse can be roughly seen as an 
`imperfect' ellipsoidal model, where the virialization is 
identified with the collapse of the second axis.

Figure \ref{ellipsoid_evolution_approx_fig} shows how well the
approximation works in the LCDM model assumed in this study. 
Solid lines show numerical solutions of equation~(\ref{ellipsoid_evolution})
in comoving units; dashed-dotted lines show the analytic approximation 
(equation~\ref{WS_approx}).
As evident from the figure, the approximation is rather good until 
the collapse of the third axis, as expected.


\subsection{The axis ratio plane: dark halo evolution} \label{axis_ratio_plane}

The evolution of triaxial objects can be conveniently described by 
a two-dimensional `axis ratio plane', showing the shortest-to-longest 
($A_1/A_3$) versus intermediate-to-longest ($A_2/A_3$) axis ratios.  
Figure \ref{ax_rat_plane} shows what our ellipsoidal collapse model 
predicts for a few different combinations of prolateness and ellipticity, 
as indicated in the panels, when 
 $\delta(t_{\rm i})= [D(t_{\rm i})/D(t_{\rm 0})] \delta(t_0)$ and $z_{\rm i}=39$. 
Note that $\delta(t_0)=1.6753$ is again the usual spherical collapse 
linear value in the concordance cosmology, at the present time. 

The axis ratio plane provides useful insights into the evolution of 
dark matter halos. The trajectory of a collapsing object in the plane 
is as follows:  down and to the left, as the shortest axis shrinks
more rapidly than the other two (first line); then further to the left
and slightly upwards, after the shortest axis has frozen out, while 
the second axis continues to shrink faster than the longest axis 
(second line); finally upwards and to the right, after the second 
axis has also frozen out, so only the longest is shrinking (third 
line), until the third axis also freezes out, at which point the object 
is defined as being virialized. 

The slope of the first line is steeper when $p=e/2$, shallower 
when $p=-e/2$.  The evolution proceeds further down the plane as 
$e$ increases, at fixed $p$, and is confined in the region between 
the bisector of the plane and the line characterized by $p=e/2$. 
Hence, when $p=e/2$ and $e > 0$, the object is more likely to form 
a pancake (i.e. lower right region of the plane), while when $p=-e/2$ 
a filament is more likely to occur (i.e. lower left region).
Also, for small values of ellipticity, the final shape does not depart
significantly from the spherical case (upper right region of the plane).   

The fact that $p$ determines the slope of the initial motion in the 
plane, whereas $e$ controls how far the initial collapse progresses, 
can be understood as follows.  Initially, i.e., when $\lambda_1 \ll 1$, 
\begin{equation}
\frac{A_1}{A_3} = \frac{1-\lambda_1}{1-\lambda_3} \simeq 1 - 2\delta(t_{\rm i}) e,
\label{ax_rat_par_one}
\end{equation}
\begin{equation}
\frac{A_2}{A_3} = \frac{1-\lambda_2}{1-\lambda_3}
              \simeq \frac{A_1}{A_3} + \delta(t_{\rm i})(e+p).
\label{ax_rat_par_two}
\end{equation}
Therefore, for a given $\delta(t_{\rm i})$, the ratio $A_1/A_3$ is a 
function of $e$ only: if $e$ increases then $A_1/A_3$ decreases,   
and at fixed $A_1/A_3$ (or, equivalently, for a given $e$) the prolateness
$p$ increases as $A_2/A_3$ increases. This effect is indicated in Figure
\ref{ax_rat_plane} by the circular arrows.

To next order in $\delta(t_{\rm i})$ 
\begin{eqnarray}
\label{proxy_ws_2nd}
\frac{A_{1}(t)}{A_{3}(t)} &\simeq& \frac{A_2(t)}{A_3(t)} 
      \frac{1 - \delta(t_{\rm i})(1+3e+p)/3}{1-\delta(t_{\rm i})(1-2p)/3}
      \times \nonumber \\
&& \Big \{ 1 - {\delta(t_{\rm i}) \over 3} {a(t) \over a_{\rm e}(t)}
   (3e + p) \Big [1+{D(t) \over D(t_{\rm i})} - {a_{\rm e} (t) \over a(t)}
  \Big ] \Big \}            \over  \Big \{ 1 + {2 \delta(t_{\rm i})
  \over 3} {a(t) \over a_{\rm e}(t)} p \Big [1+{D(t) \over D(t_{\rm
  i})} - {a_{\rm e} (t) \over a(t)} \Big ] \Big \}. 
\end{eqnarray}
In particular, when $p=0$:
\begin{equation}
\frac{A_{1}(t)}{A_{3}(t)}
  \simeq (1+3e) \Big \{ 1 + \delta(t_{\rm i}) e \Big [1 - \Big (1+ {D(t)
  \over D(t_{\rm i})}
  \Big ) {a(t) \over a_{\rm e}(t)} \Big ] \Big \} {A_2(t) \over A_3(t)}.
\label{proxy_ws}
\end{equation}

Motivated by this, we have parametrized the evolution in this plane 
as follows. If $f(e)$ denotes the distance along the first line, then, 
for a given $p$, 
\begin{equation}
 f(e) = \sqrt{(1 - A_2/A_3)^2 + (1 - A_1/A_3)^2},
\label{parametrization_one}
\end{equation}
\noindent and
\begin{equation}
 \frac{A_1}{A_3} - k(p) \frac{A_2}{A_3} = b.
\label{parametrization_two}
\end{equation}
The constraint that $A_1/A_3=1$ when $A_2/A_3=1$ implies that
$b=1-k(p)$, and therefore:
\begin{equation}
f(e) = \Big(1-\frac{A_2}{A_3}\Big) \sqrt{1+k^2(p)}.
\label{parametrization_three}
\end{equation}  
We found $k(p) = 1.858$ when $p=0$,
$k(p) = 3.803$ when $p=e/2$, and $k(p) = 1.253$ when $p=-e/2$ for the
collapse of the shortest axis.
For the turnaround of the second line, parametrized in the same 
fashion, $k(p) = 1.084$ when $p=0$, $k(p) = 1.253$ when $p=e/2$ 
and $k(p) = 1.034$ at $p=-e/2$.
The slope of the final part of the evolution (third line), 
when $A_2$ and $A_1$ are both frozen, is simply given by 
$A_1(t_{\rm f})/A_2(t_{\rm f})$.  Hence, it approaches unity (i.e. 
parallel to the bisector of the plane) when $A_1/A_3 \simeq A_2/A_3$, 
and it is always lesser than unity otherwise, since 
$A_1(t_{\rm f}) \le A_2(t_{\rm f})$.



\section{Initial and evolved halo shapes}\label{initial_conditions}

The discussion above shows how the initial values of $\delta,e,p$ 
determine the future evolution of the object, including its shape.  
Our next task is to determine how the initial distribution of $e$ and 
$p$ values depends on halo mass.  We use the algorithm of 
Sheth \& Tormen (2002) to do this.  Briefly, this requires the generation 
of the joint distribution of $\delta,e,p$ as a function of smoothing 
scale $R_{\rm i}$, finding the largest scale at which 
$\delta \ge \delta_{\rm ec}(e,p)$, and then associating the values 
$(\delta_{\rm ec},e,p)$ to the patch from which a halo of mass 
$M = \bar\rho\,4\pi R_{\rm i}^3/3$ formed. 
Since this algorithm has been used by other authors since 
(Sandvik et al. 2007), we do not provide details here, but refer the 
reader to Appendix~B in Sheth \& Tormen (2002), and to 
Bond et al. (1991), Bower (1991) and Lacey \& Cole (1993) for further details.  


\subsection{Mass-dependent shapes from scale-dependent initial conditions}

\begin{figure}
\begin{center}
\includegraphics[angle=0,width=0.48\textwidth]{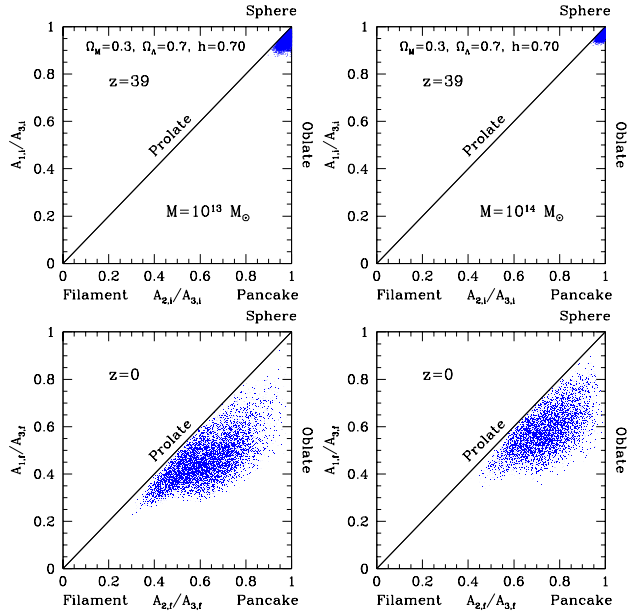}  
\caption{Distribution of initial (top panels) and final (bottom panels) 
         axis ratios in our model, for halos of mass 
         $M=10^{13}h^{-1} M_{\odot}$ (left) and $10^{14} h^{-1}M_{\odot}$ (right). 
         Initial conditions are specified by combining the statistics 
         of Gaussian random fields with the ellipsoidal collapse model, 
         following Sheth \& Tormen (2002).}
\label{in_fin_shapes}
\end{center}
\end{figure}

Figure \ref{in_fin_shapes} illustrates how our model for halo shapes
works.  The top left panel shows the distribution of initial axis 
ratios for small patches ($M=10^{13} h^{-1}M_{\odot}$) which the ellipsoidal 
collapse model predicts are destined to become low mass halos at the 
present time.  
The bottom left panel shows the distribution of final axis ratios at 
virialization (see Section \ref{nl_dynamics}). 
The panels on the right show the analogous quantities for more massive 
halos ($M=10^{14} h^{-1}M_{\odot}$). 

The plot shows that halos are, in general, predicted to be triaxial, 
with a slight tendency to be more prolate than oblate.
In addition, massive halos are predicted to be more spherical than 
less massive halos, both initially and finally. This is primarily a 
consequence of the fact that the larger patches from which massive 
halos form have, on average, smaller values of ($e,p$). The trend 
is consistent with the findings of Bernardeau (1994), who argued that 
in perturbation theory larger halos are expected to be rounder.  

\begin{figure*}
\centering  
\includegraphics[angle=0,width=0.95\textwidth]{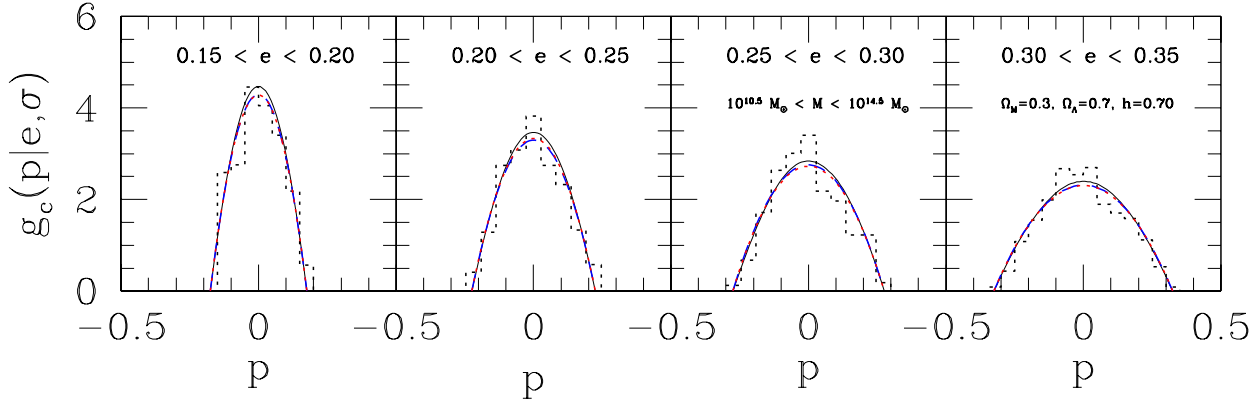}
\caption{Distribution of prolateness at fixed halo mass and ellipticity,
  $g_{c}(p|e,\sigma)$, in our model (dotted histograms). Theoretical curves
  (solid lines) are from equation (\ref{p_given_e}) with different
  values of $e$, as specified in each panel. The initial distribution
  $g(p|e,\sigma, \delta)$ is also shown, 
  with long-dashed lines (i.e. equation \ref{p_given_e_no_barrier}),
  as well as its approximation (i.e. dotted lines). See the
  main text for more details.}
\label{pro_given_eli} 
\end{figure*}


\subsection{The conditional prolateness distribution}  \label{prolateness}

The joint distribution of $\lambda_1 \ge \lambda_2 \ge \lambda_3$ was 
derived by Doroshkevich (1970) for Gaussian random fields (see 
Lam, Sheth \& Desjacques 2009 for an extension to non-Gaussian fields 
that are of the `local' type).  The equivalent expression, in terms 
of e, p and $\delta$, is:
{\setlength\arraycolsep{3pt}
\begin{eqnarray}
\label{epdelta}
\lefteqn{g(e,p,\delta|\sigma) = p(\delta|\sigma) \, g(e, p|\delta,\sigma)
{}} \nonumber \\
&=& {} \frac{e^{-\frac{\delta^2}{2 \sigma^2}}}{\sigma \sqrt{2\pi}}
 \frac{1125}{\sqrt{10 \pi}} e (e^2 - p^2)
 \Big(\frac{\delta}{\sigma}\Big)^5 e^{-\frac{5}{2}\frac{\delta^2}{\sigma^2} (3 e^2 + p^2)}.
\end{eqnarray}}
Doroshkevich's formula 
is the product of two independent distributions, a Gaussian for 
$\delta/\sigma$, and one which is a combination of the other five 
independent elements of the deformation tensor of which the 
$\lambda_{\rm j}$ are the eigenvalues (Sheth \& Tormen 2002).  
Integration of equation~(\ref{epdelta}) over $\delta$ yields the joint
distribution of ellipticity and prolateness, $g(e,p|\sigma)$.  
A further integration over $p$ gives the distribution of ellipticities 
as a function of halo mass ($\sigma$), $f(e|\sigma)$.  This integral 
can be computed analytically (c.f. equation 24 of Lam et al. 2009).  

For a given ellipticity, we can predict the distribution of $p$'s 
that will produce collapsed objects at any given redshift 
(say, $z = 0$).  In detail, for $\delta_{\rm ec}(e,p)$ as in 
equation~(\ref{deltaec_approx}), we define
 $\Theta (\delta, e, p) = 1$ if $\delta \ge \delta_{ec}(e,p)$ and 
 $\Theta (\delta, e, p) = 0$ otherwise.    
Then we impose the requirement that the perturbation had collapsed 
(at $z=0$) by computing 
{\setlength\arraycolsep{3pt}
\begin{eqnarray}
\label{p_given_e}
\lefteqn{g_{\rm c}(p|e,\sigma) 
 = \frac{g_{\rm c}(e,p|\sigma)}{\int_{-e}^{e} {\rm d}p~g_{\rm c}(e,p|\sigma)}
    {}} \nonumber \\
&=& {} \frac{\int_{0}^{\infty} {\rm d} \delta\,g(e,p|\delta,\sigma)\,
    p(\delta|\sigma) \ \Theta [\delta \ge \delta_{\rm ec} (e,p)]}
   {\int_{\rm -e}^{\rm e} {\rm d}p \int_{0}^{\infty} 
    {\rm d}\delta\, g(e,p|\delta,\sigma)\,
    p(\delta|\sigma)\, \Theta [\delta \ge \delta_{\rm ec} (e,p)]}.
\end{eqnarray}}
Plots of equation~(\ref{p_given_e}) are shown in 
Figure~\ref{pro_given_eli} with solid lines.  
We can further quantify if the difference between $g(p|e,\delta,\sigma)$
and $g_{\rm c}(p|e,\sigma)$ is relevant (i.e. dependence of barrier
and epoch) by computing:
{\setlength\arraycolsep{3pt}
\begin{eqnarray}
\label{p_given_e_no_barrier}
\lefteqn{g(p|e,\delta,\sigma) =
  \frac{g(e,p|\delta,\sigma)}{\int_{-e}^e {\rm d}p~g(e,p|\delta,\sigma)}  {}} \nonumber \\ 
&=& {} {25y^3\,(1-p^2/e^2)\, {\rm e}^{(5y^2/2)(1-p^2/e^2)}/e  \over  
        \sqrt{10\pi}(5y^2-1)\,{\rm e}^{(5y^2/2)} 
        {\rm erf}(\sqrt{5y^2/2}) +10y} .
\end{eqnarray}}
\noindent where $y = e\delta/\sigma$.
Long-dashed lines in Figure~\ref{pro_given_eli} show this distribution: 
as evident from the panels, the difference between (\ref{p_given_e}) and
(\ref{p_given_e_no_barrier}) is negligible. In fact, 
if one ignores the exponential prefactor in equation (\ref{epdelta}), 
then it is straightforward to show that 
\begin{equation}
 g(p|e,\delta,\sigma) \simeq g(p|e) = (3/4)(1-p^2/e^2)/e. 
\end{equation}
Dotted curves in Figure~\ref{pro_given_eli} show that this is a very good 
approximation.  


\subsection{Universal conditional axis ratio distributions at late times} \label{universal}

In equation~(\ref{p_given_e}), the mass and epoch dependencies actually 
cancel out, i.e. $g_{\rm c}(p|e,\sigma) \equiv g_{\rm c}(p|e)$. 
This suggests that nonlinear evolution will not change the conditional 
distribution $g_{\rm c}(p|e,\sigma)$ from that which is set by the 
initial conditions.  
To see why, recall that the axis ratio $A_1/A_3$ is specified once 
$\delta(t{\rm _i})$ and the ellipticity are fixed 
(c.f. equation~\ref{ax_rat_par_one}).  
Moreover, if $\delta(t{\rm _i})$ and $e$ are fixed, then 
$A_1/A_2 = (A_1/A_3)/(A_2/A_3)$ is only a function of the 
prolateness (c.f. equation~\ref{ax_rat_par_two}). 
Hence, for a fixed mass, equation~(\ref{p_given_e}) can be
equally expressed in terms of ($e, p$), or as a function of 
($A_1/A_2, A_1/A_3$).  I.e., if we define 
$A_{12}\equiv A_1/A_2$ and $A_{13}\equiv A_1/A_3$, then 
\begin{eqnarray}
\label{js_explain}
p(A_{12}|A_{13},\delta,\sigma)
  &=& \frac{g(A_{12},A_{13}|\delta,\sigma)}{p(A_{13}|\delta,\sigma)}\nonumber\\
  &=& \frac{3}{2 (1-A_{13})} \, 
       \Big [1 - \frac{(2A_{12}-1-A_{13})^2}{(1-A_{13})^2} \Big ] \nonumber \\ 
   &\times& \exp \Big \{ - {5 \over 8 \sigma^2} (2A_{12} -1 -A_{13})^2 \Big \},
\end{eqnarray}
where the final expression follows from setting 
\begin{equation}
 g(A_{12},A_{13}|\delta,\sigma)\,{\rm d}A_{12}\,{\rm d}A_{13} 
 = g(e,p|\delta,\sigma)\,{\rm d}e\, {\rm d}p.
\end{equation}
At small masses, the exponential term $\to 1$, making this distribution 
almost universal, i.e. independent of mass and epoch.  
This remarkable feature will guide the interpretation of our results 
on halo shapes in the next section. 



\section{Comparison with high resolution simulations} \label{simulation_comparison}

The previous section showed how we combine the nonlinear ellipsoidal 
model for halo collapse and virialization (Section \ref{nl_dynamics}) 
with the excursion set theory (Section \ref{initial_conditions}), 
to model the initial and evolved shape distributions of dark matter 
halos.  Here, we investigate to what extent our simple analytic
prescription is able to reproduce results from $N$-body
simulations of Jing \& Suto (2002).  

For their statistical analysis of halo shapes, Jing \& Suto (2002) 
identified halos at $z=0, 0.5, 1$ in simulations of a cosmology with 
 ($\Omega_{\rm M} =0.3, \Omega_{\rm \Lambda} = 0.7, h=0.7$).
The simulations used $N=512^3$ particles in a 100 h$^{-1}$Mpc box. 
Halos were identified using a friends-of-friends (FOF) algorithm with 
link-length $b=0.1d$, where $d$ is the interparticle separation.
Each of their FOF halos contains more than $10^4$ particles.  
The lower mass limit of their halo catalog is 
$6.2 \times 10^{12}h^{-1}M_\odot$.  
They provided a set of fitting formulae which allow simple tests of 
our triaxial model:  
in particular the mass and redshift dependence of the axis ratios, 
and the probability distribution functions of the axis ratios. 

\begin{figure}
\centering
\includegraphics[angle=0,width=0.48\textwidth]{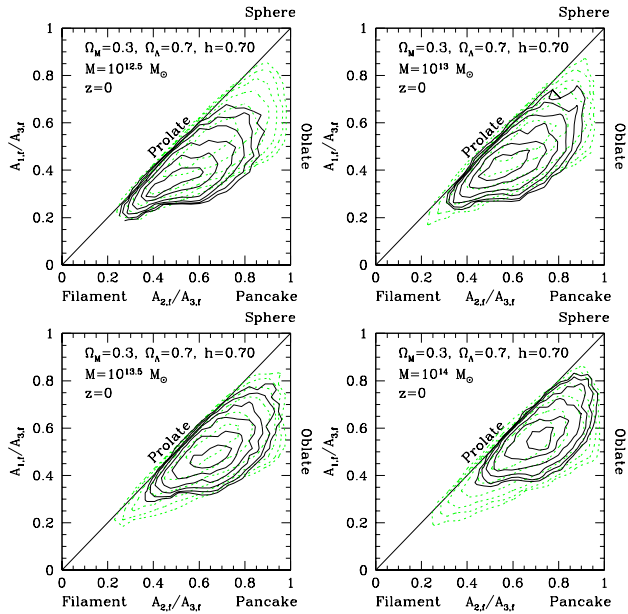}  
\caption{Distribution of final axis ratios in our model for halos of
         different masses (solid contours), as specified in the
         panels. Each halo has been evolved using the ellipsoidal 
         collapse model. The generic evolution is initially towards 
         an oblate, pancake-like structure, followed by a
         shift towards a more prolate shape, as the other axes also
         begin to shrink. Dotted lines are
         the joint distribution which Jing \& Suto (2002) report
         describe the halos in their simulations, in the concordance
         cosmology ($\Omega_{\rm M} =0.3,
         \Omega_{\rm \Lambda} = 0.7, h=0.7$).}
\label{shapes_in_sim}
\end{figure}  


\subsection{Joint distribution of axis ratios}

By averaging over the correct, mass dependent, distribution of initial 
shape parameters, each evolved through the ellipsoidal collapse model, 
we are also able to make predictions for how the distribution of final 
shapes depends on halo mass. The solid contours in Figure \ref{shapes_in_sim} 
show the joint distribution of ($A_2/A_3, A_1/A_3$) we predict; contours show 
levels where the probability has fallen to $1/2, 1/4, 1/8, 1/16$ of the 
maximum value. The dotted lines represent the joint distribution which 
Jing \& Suto (2002) report describe the halos in their simulations. 
Note, for the moment, that our model appears to be in reasonable 
agreement with the simulations only around 
$M \simeq 10^{\rm 13} h^{-1}M_{\odot} \simeq M_{*}$ .
While this agreement is gratifying, 
the fact that the mismatch is mass-dependent is potentially troubling.
Jing \& Suto find that the low mass halos are actually slightly 
more spherical than massive halos.  Although the effect is weak, 
it has since been confirmed by a long list of authors 
(see Section \ref{intro}).  In contrast, in our model, it is the 
massive halos which are expected to be more spherical 
(Figure~\ref{in_fin_shapes}).  The trend we find is consistent with 
perturbation theory (Bernardeau 1994; Pogosyan et al. 1998; see Park et al. 2007, 
Dalal et al. 2008 and Park, Kim \& Park 2010 for more recent 
discussion of this point).  

\begin{figure*}
\centering 
\includegraphics[angle=0,width=0.80\textwidth]{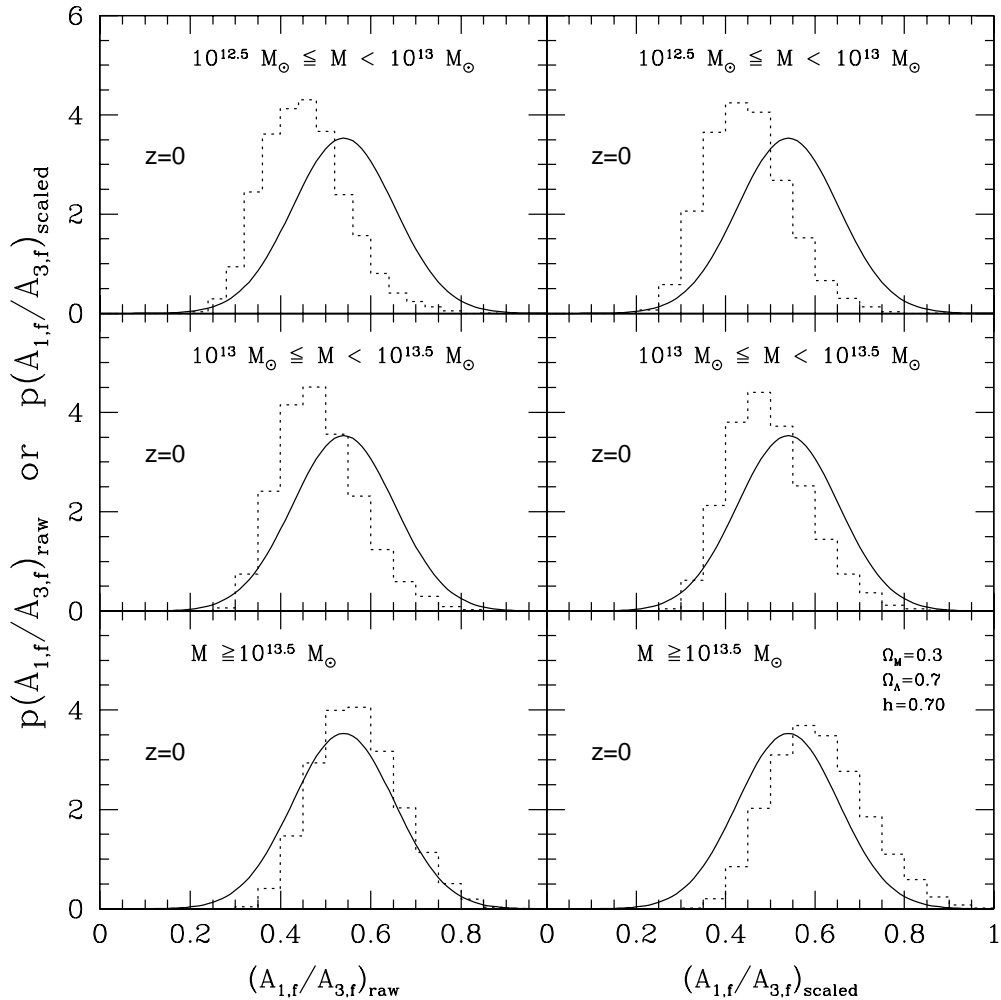}  
\caption{Distribution of the minor-to-major axis ratio, $A_1/A_3$, 
         in our model (dotted) and in Jing \& Suto's (2002) numerical 
         simulations (solid). Raw data are plotted in the left panels,
         while in the right panels data are rescaled according to their 
         prescription.}
\label{JS_comparison}
\end{figure*}

To show this more clearly, Figure \ref{JS_comparison} compares the 
final distribution of $A_1/A_3$ in our model (dotted) with the 
distribution in Jing \& Suto's simulations (solid) at $z=0$. 
Again, note that the agreement is reasonable around $M_{*}$, but that 
the predicted distribution is broader than the simulations at low and 
high masses. Note also that the left panels in the figure show our 
results without any arbitrary rescaling, while in the right panels we 
apply the empirical rescaling they suggest, which effectively removes 
the trend with mass.  

Park, Kim \& Park (2010) have argued that the main reason for this
discrepancy may be due to the halo definition itself, and to the halo
environment.  They used a more sophisticated halo finding algorithm (from Kim \& Park 2006) 
which identifies gravitationally self-bound and tidally stable halos.  
They showed that the dependence of $A_1/A_3$ on local density is stronger 
for more massive \textit{isolated} halos, and argued that tidal 
interactions between neighboring halos make them more spherical on 
average. In particular, Park, Kim \& Park (2010) 
pointed out that the main problem
relies primarily on the FOF halo finder.
This is because the FOF identification
scheme cannot distinguish between objects connected by
filaments. For example, if two massive and almost spherical neighboring halos are slightly
connected by a filament, the FOF halo finder will consider those objects as one big
structure, and erroneously assign a rather elongated shape to it. 
Therefore, the FOF scheme will find that more massive objects are more
elongated, while in reality this is an artifact of the FOF procedure. 
Park et al. (2010) went one step further; after running their
halo finder algorithm, they identified subhalos within FOF halos, and
classified their FOF halos
into central (i.e. the most massive halos in each group of halos
and subhalos), satellite (i.e. halos which are not the central ones in each group of
halos and subhalos), and isolated halos (i.e. structures which do not
overlap with other halos, but which can have close neighbor halos and subhalos).  
After this classification, they found that the degree of halo sphericity increases with
mass (i.e. more massive objects are rounder) for \textit{isolated}
halos, up to about $2 \times 10^{14} h^{-1} M_{\odot}$ -- a similar trend we find in our theoretical study.

\subsection{Conditional axis-ratio distributions} \label{conditional_distributions}

\begin{figure*}
\includegraphics[angle=0,width=0.45\textwidth]{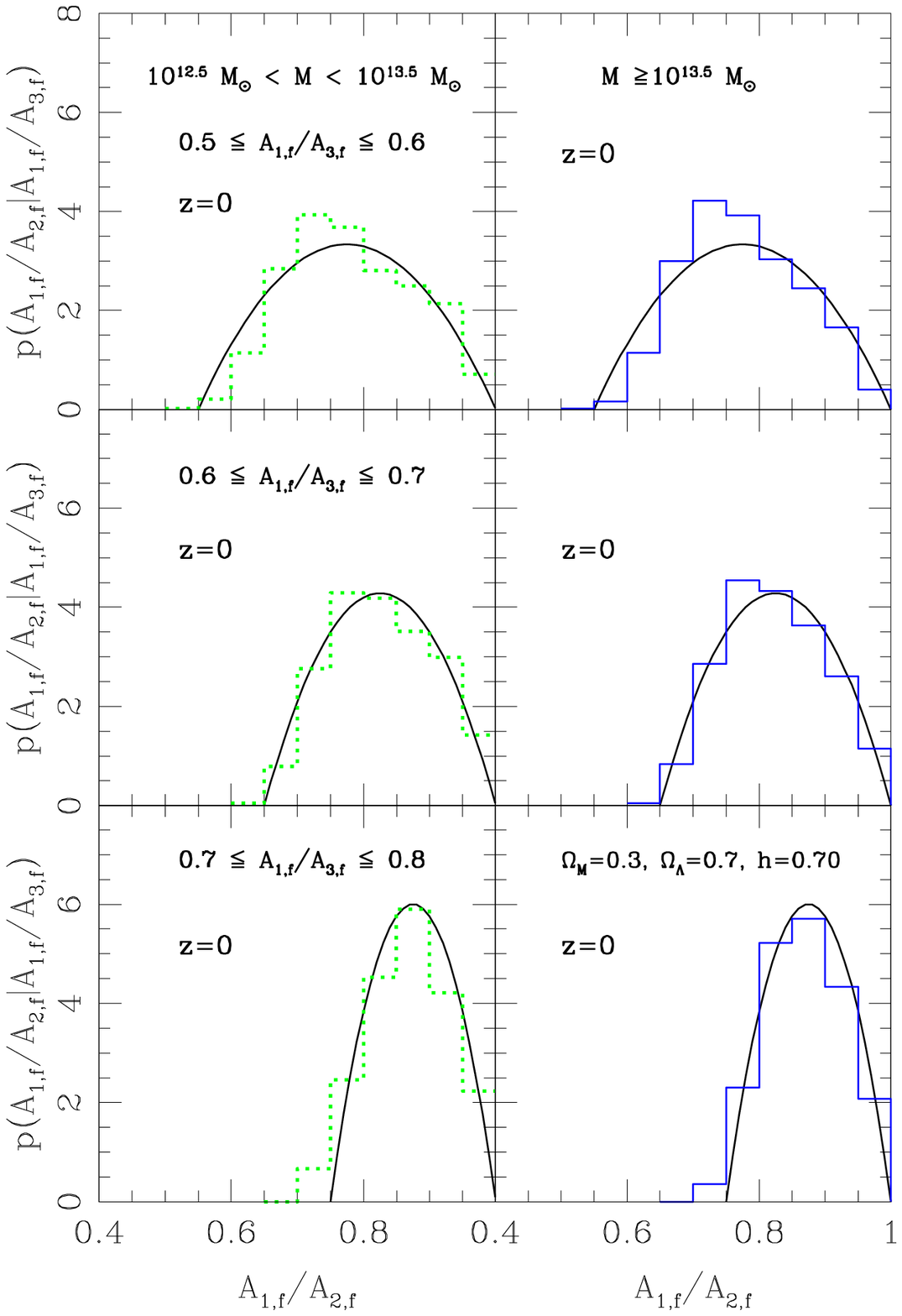}
\includegraphics[angle=0,width=0.45\textwidth]{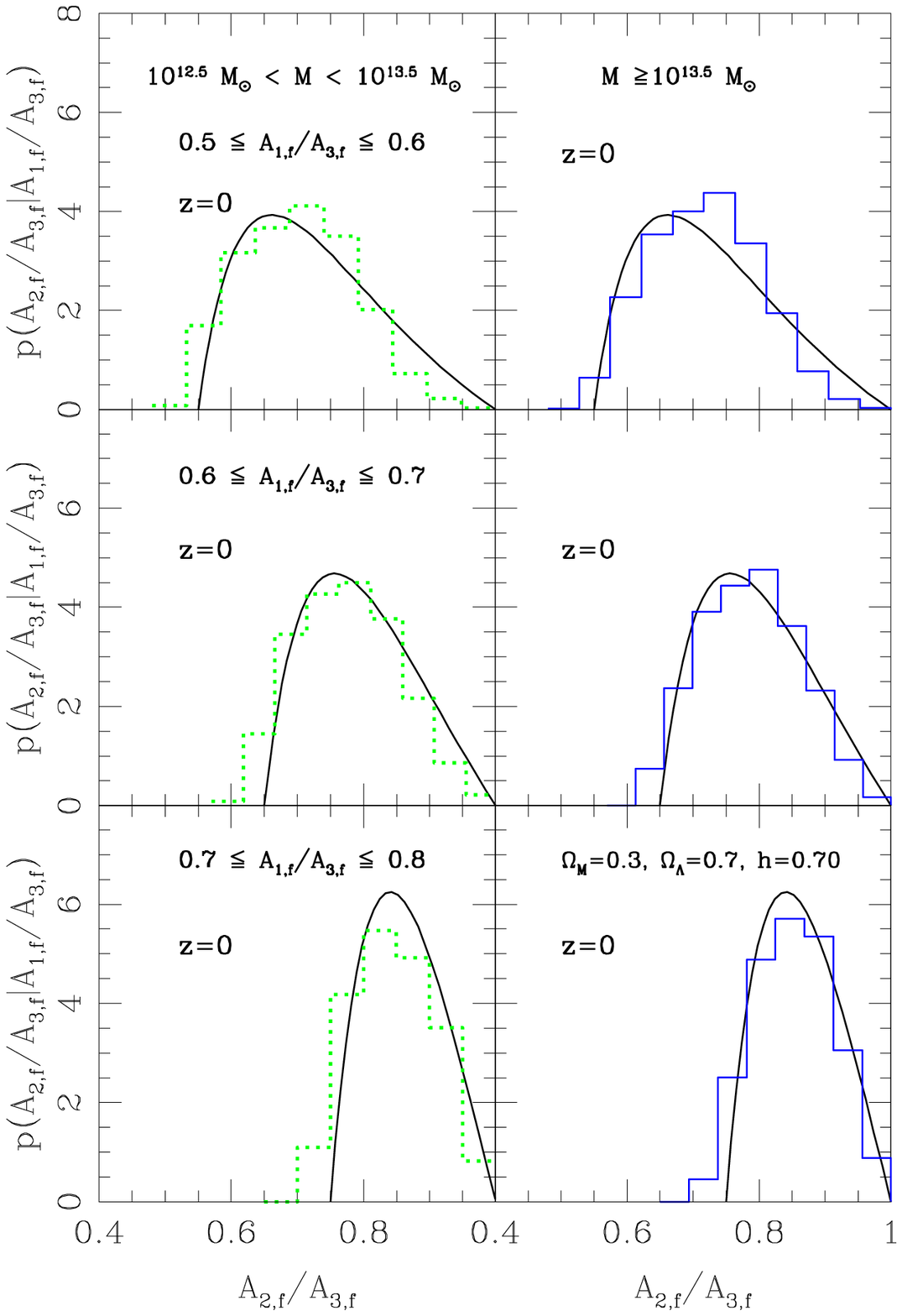}
\caption{Conditional distributions ($A_1/A_2|A_1/A_3$) -- left panel
         -- or ($A_2/A_3|A_1/A_3$) -- right panel -- in 
         our model (histograms), and in the numerical simulations of 
         Jing \& Suto (2002) (solid lines).  Results are shown 
         for two different mass ranges (both at $z=0$) and for different 
         values of $A_1/A_3$, as labeled. }
\label{p_cbca} 
\end{figure*}

The left panel of Figure \ref{p_cbca} compares the conditional minor-to-intermediate
axis distribution of $A_1/A_2$ for a given range of $A_1/A_3$ in our model 
(histograms), with Jing \& Suto's simulations (solid lines) at $z=0$.
In this case, instead, we find remarkably good agreement with the 
numerical measurements.

Recall that we had argued that the conditional prolateness 
distribution $g(p|e,\delta,\sigma)$ in the initial conditions 
(equation~\ref{js_explain} and Figure~\ref{pro_given_eli}), depends 
only weakly on $\delta/\sigma$, so we expect the conditional 
minor-to-intermediate axis ratio distribution of the evolved object, 
$p(A_{12}|A_{13},\delta, \sigma)$, to be similarly universal -- 
independent of mass and time; whether this equivalence is
due to statistics, rather than physics, is subject of ongoing work. 
The agreement with Jing \& Suto's 
simulations suggests that $p(A_{12}|A_{13},\delta, \sigma)$ is indeed 
preserved from the initial conditions.  In fact, except for the exponential 
factor in our equation~(\ref{js_explain}), our theoretically motivated 
expression for this distribution is {\em exactly} the same as that of 
Jing \& Suto's empirical fitting formula.

More recently, Wang et al. (2010) have also found that, in 
their simulations, $p(A_{12}|A_{13})$ is independent of mass and epoch. 
They also noted that the final redshift dependence of the 
short-to-intermediate axial ratio is much weaker than the other two 
ratios, indicating that new material tends to be accreted along the 
major axes of halos.  Our formula~(\ref{js_explain}) provides the 
theoretical/physical explanation for their findings.  

In this context, we note that Lee, Jing \& Suto (2005) also presented 
an analytic expression for the axis ratio distribution of triaxial 
objects, and claimed that it successfully reproduces the conditional 
intermediate-to-major axis ratio distribution.  However, in their 
Figure~5 which is supposed to support this claim, it is the 
\textit{minor-to-intermediate} distribution from the numerical 
simulations, $p(A_1/A_2|A_1/A_3)$, which is compared with their theory 
curve for the \textit{intermediate-to-major} axis ratio distribution, 
$p(A_2/A_3|A_1/A_3)$.  While this invalidates their claim (see also
Appendix \ref{ljs_problems} for more discussion on the flaw in their model), it does 
raise the question of why the formula for one distribution happened 
to describe another?  

Our model shows why.  
When $\lambda_1 \ll 1$, then
 $A_1/A_2 \simeq 1 - \delta(t_{\rm i})(e+p)$, while
 $A_2/A_3 \simeq 1 - \delta(t_{\rm i})(e-p)$. 
However, the sign difference in the prolateness does not alter
the Jacobian of the two different transformations:  
($e,p \rightarrow A_2/A_3,A_1/A_3$) and 
($e,p \rightarrow A_1/A_2, A_1/A_3$).  
Hence,
 $p(A_2/A_3|A_1/A_3)\, {\rm d}(A_2/A_3) \equiv
  p(A_1/A_2|A_1/A_3)\,{\rm d}(A_1/A_2)$, and therefore
\begin{equation}
p(A_2/A_3|A_1/A_3) = {(A_1/A_2) \over (A_2/A_3)} 
\cdot p(A_1/A_2|A_1/A_3).
\end{equation}
Thus, at fixed $A_1/A_3$, the minor-to-intermediate or the 
intermediate-to-major axis ratios distributions are both 
equivalent, and so `universal'.
Although \textit{equivalent}, clearly these distributions are indeed
different and neatly distinguishable:
the right panel of Figure \ref{p_cbca} compares now the conditional intermediate-to-major
axis distribution of $A_2/A_3$ for a given range of $A_1/A_3$ in our model 
(histograms), with Jing \& Suto's simulations (solid lines) at $z=0$.
As evident from the plots,  $p(A_2/A_3|A_1/A_3)$ is more asymmetric than
$p(A_1/A_2|A_1/A_3)$, and the difference is more significant for lower
values of $A_1/A_3$.

\subsection{Caveats}

In general, a direct comparison between our theoretical model and 
$N$-body simulations is both challenging and subtle. This is primarily 
because there is still no unanimous agreement about how to define a
gravitationally bound object; i.e the very definition of what a 
non-spherical dark matter halo is, both in simulations and observationally,
is a non-trivial problem.  Determining which material belongs to a 
halo and what lies beyond is an open problem (Prada et al. 2006; 
Bett et al. 2007; Diemand, Kuhlen \& Madau 2007; Cuesta et al. 2008).
Common methods for defining virialized halos in simulations include: 
spherical overdensity, tree algorithms based on the branches of the 
halo merger trees, two-step procedures with post-processing, 
maximum circular velocity, etc., 
(see Klypin et al. 2010, and references therein for more discussion).
Clearly, the distribution of halo shapes measured from simulations
depends critically on the halo definition.  For instance, 
Bett et al. (2007) found that spherical overdensity halos are more 
spherical than FOF or TREE halos, and that FOF halos show a 
much broader distribution of shapes and a strong preference for 
prolateness.  This fact complicates the comparison and interpretation 
of theory against any numerical study.  

Since we are most interested in halo shapes, measurements which 
do not perform spherical averages are more closely related to 
our models.  Jing \& Suto (2002) used one such method:  a 
friend-of-friend algorithm (Davis et al. 1985; 
Lacey \& Cole 1994).  This is a percolation scheme that links together 
all particles that are closer than $b=0.2d$, where $d$ is the mean 
interparticle separation.  This value of $b$ returns objects which 
are approximately 200 times the background density:  it makes 
no assumptions about the shape of the resulting object.  
However, it may group distinct halos together into the same object, 
confusing the comparison with theory (White 2001; Tinker et al. 2008; 
Lukic et al. 2009).  This problem is more serious for halos in higher 
density regions; since high density regions have top-heavy mass 
functions (Sheth \& Tormen 2002), massive halos identified by this 
algorithm may be erroneously considered to be elongated.  

Jing \& Suto (2002) used $b=0.1d$, so their FOF clumps are smaller 
and denser than those defined using the more conventional $0.2d$.  
As a result, they are less likely to suffer from spuriously linked 
halos.  However, in effect, their procedure identifies only the central 
parts of the more conventional halo.  

As a further complication, having settled on a halo definition, some 
authors only study shapes of a subset of the halo population.  
E.g., Jing \& Suto (2002), only analyzed halos which looked `relaxed': 
at fixed mass, this is almost certainly a more homogeneous subset of 
the entire population.  Our model does not account for this additional 
selection.  

Despite such difficulties in the model/simulation comparison, 
we believe such an exercise is still useful, because there are other 
reasons why we anticipate disagreement between model and simulations.  
E.g., the ellipsoidal collapse itself is a significant oversimplification 
of the more complex nonlinear gravitational evolution -- it cannot be 
expected to fully capture the dynamics of $N$-body simulations.  
Despite all these caveats, we believe the remarkably good agreement between the 
measured conditional axis ratio distributions and our 
equation~(\ref{js_explain}) (Figure~\ref{p_cbca}) is non-trivial.  



\section{Discussion} \label{discussion}


We presented a model to describe the shape distribution of bound objects
based on Rossi (2008), which is a simple extension of that introduced by 
Bond \& Myers (1996) and used by Sheth, Mo \& Tormen (2001) to estimate 
the abundance by mass of these objects.
Our analytic prescription is made of two independent parts: 
one is a scheme for how an initially spherical patch evolves and
virializes, for which we adopted the ellipsoidal collapse dynamics
(Section \ref{nl_dynamics}, Figure \ref{ellipsoid_evolution_fig});  
the other is the correct assignment of initial shapes to halos of 
different masses, achieved through the excursion set formalism 
(Section \ref{initial_conditions}, Figure \ref{in_fin_shapes}).  
Along the way, we discussed an analytic approximation to the evolution,
considerably more rigorous than the Zeldovich approximation (Section
\ref{analytic_proxy}, Figure \ref{ellipsoid_evolution_approx_fig}). 
And we introduced a useful planar representation of halo axis ratios, 
which gives a simple intuitive framework for discussing the dynamics 
of collapsing regions (Section \ref{axis_ratio_plane}, 
Figure \ref{ax_rat_plane}).

We showed that the model is able to provide a reasonable description 
of the minor-to-major axis ratio distribution seen by 
Jing \& Suto (2002) in their numerical simulations only within a 
limited mass range (i.e. around $M\simeq 10^{\rm 13}
h^{-1}M_{\odot} \simeq M_{*}$).  Outside this range, Jing \& Suto found 
that low-mass halos are preferentially more spherical than high-mass 
halos, whereas in our model, the high-mass halos are more spherical 
(Figures \ref{shapes_in_sim} and \ref{JS_comparison}).  
We argued that some the disagreements may originate from differences 
in halo types and mass range, and may be alleviated if we consider 
only \textit{isolated} halos.  This is because Park, Kim \& Park (2010) find 
that, on average, high-mass isolated halos are indeed more spherical 
than low-mass ones, and the difference is larger in high density/shear 
regions (see their Section 3.4 and their Figures 9, 10 and 11, for
details on how to select isolated clusters from simulations).
Note also that isolated clusters can be readily identified from an
observational catalog, as done for example in Park \& Hwang (2009) for a
sample of Abell galaxy clusters (their Section 2.2 gives a detailed
explanation on how the selection is performed).
The finding of Park, Kim \& Park (2010) has
an interesting connection to recent observational work on 
mass scales where the hypothesis that a halo hosts only one galaxy 
may be appropriate.  For early-type galaxies in the SDSS, although 
the observed projected axis ratio $b/a$ increases with increasing 
stellar mass or luminosity, it decreases at the highest masses 
(Bernardi et al. 2008, 2010).  This reversal is thought to arise 
because the most massive galaxies tend to be in regions (e.g., at 
cluster centers) where recent radial mergers may have made them prolate.  
Fasano et al. (2010) come to a similar conclusion based on an analysis 
of the intrinsic shape distribution of BCGs from the WINGS survey 
(Fasano et al. 2006; Varela et al. 2009).  They propose that 
the prolateness of the BCGs (in particular the cDs) could reflect the 
shape of the associated dark matter halos (also see Smith et al. 2010).  
Hence, while it is tempting 
to conclude that the discrepancy between mass dependent trends in 
simulations and theory will be alleviated if we only consider isolated 
halos, we note that Jing \& Suto (2002) eliminated halos in their 
simulations which were not relaxed.  If these had suffered recent 
mergers, then they may already have removed the most prolate halos from 
their sample.  Further investigation of this point is the subject of 
ongoing work.  

Despite this disagreement about the distribution of minor-to-major 
axis ratio, we found very good agreement between our model and the 
conditional minor-to-intermediate axis ratio distribution measured
from simulations (Figure \ref{p_cbca}).  
In particular, we showed that our model provides physical motivation 
for the empirical fitting formula obtained in Jing \& Suto (2002) from 
numerical studies (our equation~\ref{js_explain}).  In our model, this 
distribution is closely related to the conditional prolateness 
distribution in the initial conditions (Section \ref{prolateness}, 
Figure~\ref{pro_given_eli}).  Unfortunately, results in Lam et al. (2009) 
suggest that this distribution is unlikely to be able to discriminate 
between Gaussian initial conditions and ones where the primordial 
non-Gaussianity is of the `local' type (i.e. Rossi, Chingangbam \&
Park 2010).  

We also discussed the agreement/disagreement between
theory and numerical predictions, in the context of known problems and 
limitations in the modelling to difficulties in making shape measurements 
from simulations.

We are currently studying the following improvements and extensions 
to our model:
\begin{enumerate}
\item Our model makes the extremely simple assumption that axis lengths 
      freeze-out once they have shrunk to a sufficiently small size.  
      Subsequent violent relaxation effects associated with particles
      which collapse along other axes may change this size -- something 
      our current model does not consider. For instance, massive halos 
      are expected to have assembled their mass more recently than low 
      mass halos. Therefore, relaxation effects may have had more time 
      to change the axis ratios of low mass halos than of higher mass 
      halos. If the net effect of this relaxation is to make the halos 
      more spherical than they would otherwise have been, then accounting 
      for it may help resolve the discrepancy between our model and 
      simulations. 
      Note that halos having smaller amounts of substructures tend to 
      be closer to virial equilibrium (e.g. Giocoli et al. 2010), but 
      there has been no study of whether or not haloes with abnormally 
      small sub-structure (for their mass) are rounder.  
\item Testing different collapse criteria, and other prescriptions for
      the external tidal field (e.g., Angrick \& Bartelmann 2010). 
\item Accounting for correlations between the properties of halos and
      their environment (assembly bias), as seen in $N$-body simulations 
      (Sheth \& Tormen 2004; Ragone-Figueroa \& Plionis 2007).  
      Environmental effects can have a significant impact on the
      properties of virialized halos because, in the ellipsoidal model, 
      the critical density threshold $\delta_{\rm ec}$ depends strongly 
      on the initial values of $e$ and $p$, and these distributions 
      depend on environment (Keselman \& Nusser 2007; Wang, Mo \& Jing 2007; 
      Desjacques 2008).  In our current implementation, only the 
      initial density of the surrounding environment matters.  
\item Accounting for correlations between formation times and 
      halo shapes, as seen in $N$-body simulations 
      (Ragone-Figueroa et al. 2010).  
\item Modifying the algorithm of Sheth \& Tormen (2002) to account for 
      correlated steps, following Maggiore \& Riotto (2010) and 
      De Simone et al. (2010), when deriving our initial conditions.
\item Including the effects of baryonic physics.  
      The condensation of baryons to the centers of dark matter halos 
      is known to change their shape
      (Dubinski 1994; Holley-Bockelmann et al. 2002; 
      Kazantzidis et al. 2004; Debattista et al. 2008). 
\item Estimating correlations between halo shapes. 
      This is possible because, in our model, the final shape is 
      determined by the initial conditions, and correlations in the 
      initial conditions can be quantified. Thus, in our model, 
      correlations at the present time are obtained by taking the 
      appropriate average over the initial correlations -- this average 
      being determined by the mapping from 
       $(e,p) \rightarrow (A_2/A_3, A_1/A_3)$. 
      Most published estimates of the correlation between shapes do not
      account for this mapping.   
\item Describing the shapes of subhalos as well as voids 
      (e.g., Shandarin et al. 2006; D'Aloisio \& Furlanetto
      2007; Platen et al. 2008). 
\end{enumerate}


We conclude with the observation that while numerical studies are 
indispensable for quantifying the shapes and structures of dark matter 
halos, we hope that our simple procedure may provide a complementary 
theoretical framework for understanding halo shapes, and how these 
shapes are correlated with larger-scale structures.



\section*{Acknowledgments}

We would like to thank Changbom Park for a careful reading of the
manuscript, and for many interesting 
discussions, suggestions and encouragement.  
RKS is supported in part by NSF-AST 0908241.




\appendix


\section{Comparison with the model of Lee, Jing \& Suto (2005)} \label{ljs_problems}
The main text describes a number of inconsistencies in the work of 
Lee et al. (2005).  Besides the mismatch between $p(A_1/A_2|A_1/A_3)$ 
and $p(A_2/A_3|A_1/A_3)$ (see Section~\ref{conditional_distributions}), 
we also stated in the main text that their approach does not reduce to 
the Zeldovich approximation -- despite the fact that they believe it does.  

To see why, note that their equation~(2) gives the usual expression for 
the nonlinear density in the Zeldovich approximation -- the same one 
we use.  Namely, if we use $R_{\rm i}$ to denote the initial length of 
axis $i$, and $r_{\rm i}$ its length at a later time, then their 
equation~(2) comes from setting $r_{\rm i}/R_{\rm i} = 1 - D(t) \lambda_{\rm i},$ 
where $\lambda_i$ is the eigenvalue of the initial shear tensor.  
Note that there is \textit{no square root factor} in the relation between 
$\lambda_{\rm i}$ and the Zeldovich-evolved axis length $r_{\rm i}/R_{\rm i}$. 

Lee et al. (2005) then define ($a, b, c$) to be the eigenvalues of 
the inertia tensor.  For a homogeneous ellipsoid with axis ratios 
($A, B, C$), the inertia tensor eigenvalues ($a, b, c$) are related to 
($A^2$, $B^2$, $C^2$), whereas their equation~(7) sets $a$ equal to the 
square root of $A$. I.e., they have put the square root on the wrong side 
of the equation. Therefore, their expressions will not reduce correctly to 
the Zeldovich approximation at early times. 

Note that their equation~(8) does in fact have the correct relation between 
the \textit{semi-axis lengths} and eigenvalues (because it has been taken
correctly from Equation 7.2 of Bardeen et al.  1986).  In this notation, 
$\zeta_1 = a^2$ where the $\zeta_i$ are the eigenvalues of the inertia tensor, 
but Lee et al. (2005) incorrectly call ($a, b, c$) the eigenvalues of 
inertial tensor, when they are in fact the \textit{semi-axis lengths} 
(see text above Equation 7.2 of Bardeen et al. 1986). 
However, by equation~(15), Lee et al. are referring to $b/c$ as ratios 
of \textit{axis lengths}, rather than as ratios of \textit{eigenvalues}.  
Their sloppiness about the important difference between these two gives 
rise to the following problem. If ($a, b, c$) are axis lengths, then the 
square root factor in their equation~(7) means they will not recover the 
Zeldovich approximation, which does not have a square root factor. 
If they are eigenvalues of the inertia tensor, then they have put the 
square root in the wrong place, so again, they will not recover the 
Zeldovich approximation. 

Unfortunately, Jing \& Suto (2002) are also ambiguous about just what 
it is they measured. Their equation~(5) shows ($a, b, c$) as being 
\textit{axis ratios}, whereas their text describes them as being obtained 
from diagonalizing the inertia tensor, hence \textit{eigenvalues} 
(unfortunately, they don't say explicitly which is the square-root of 
the other). 

In a subsequent study on cosmic voids, 
Lavaux \& Wandelt (2010) used some erroneous definitions from Lee et
al. (2005) but still found excellent agreement between their
analytic framework and measurements from numerical simulations. 
This is because they where ``consistently wrong'', in that their derived quantities measured
from the simulations suffer from the same flaw present in their
theory. Hence, although their ellipticity definitions are incorrect, 
the comparison theory/simulations is still fair.
We also note that the philosophy of our approach in modelling halo
shapes is similar to theirs, although the details are different (see
their Appendix B) and they matter
(i.e. in particular the use of the ellipsoidal collapse barrier as in
Equation \ref{p_given_e}, which allows one to
distinguish between
peaks and random positions). We will expand more on this important issue
in a forthcoming publication.


\label{lastpage}


\end{document}